\documentclass[epj]{svjour}

\usepackage{graphicx}
\usepackage{fancyhdr}
\def\makeheadbox{{
 }}
\def\mytitle{My title} 
\def\myauthors{My name}  
\def\mytype{My type of session}
\def\mysession{My session}

\def\mytitle{Reflection of microwave from energy deposit by X ray irradiation in rock salt} 
\def\myauthors{Masami Chiba}    
\def\mytype{Contributed Talk}    
\def\mysession{Cosmology and Astrophysics}

\begin{document}
\title{Reflection of microwave from energy deposit by X-ray irradiation in rock salt}
\subtitle{Implication of an ultra high energy salt neutrino detector to act like a radio bubble chamber}
\author{Masami Chiba\inst{1}
\thanks{\emph{Email:} chiba-masami@c.metro-u.ac.jp} %
\and Yoko Arakawa\inst{1}\and Toshio Kamijo\inst{1}\and Shunsuke Nakamura\inst{1}\and %
Yuji Shibasaki\inst{1}\and Yasuhiro Takayama\inst{1}\and Yusuke Watanabe\inst{1}\and
Fumiaki Yabuki\inst{1}\and Osamu Yasuda\inst{1}\and %
Akio Amano\inst{2}\and Yuichi Chikashige\inst{2}\and Keisuke Ibe\inst{2}\and %
Tadashi Kon\inst{2}\and Sosuke Ninomiya\inst{2}\and Yutaka Shimizu\inst{2}\and Yoshito Takeoka\inst{2}\and%
Yasuyuki Taniuchi\inst{3}\and Michiaki Utsumi\inst{3}\and Masatoshi Fujii\inst{4}
}                     
%
%
\institute{Graduate School of Science and Engineering, Tokyo Metropolitan University, %
Hachioji-shi, Tokyo, Japan
\and Faculty of Science and Technology, Seikei University, Musashino-shi, Tokyo, Japan
\and Department of Applied Science and Energy Engineering, School of Engineering, %
Tokai University, Hiratsuka-shi, Kanagawa, Japan
\and School of Medicine, Shimane University, Izumo-shi, Shimane, Japan}
%
\date{}
\abstract{
Existence of GZK neutrinos (ultra high energy neutrinos) have been justified although the flux is very low. %
A new method is desired to use a huge mass of a detector medium to detect them. %
A fundamental study of radar method was carried out to measure %
microwave reflection from electromagnetic energy deposit by X-ray irradiation %
in a small rock salt sample. %
The reflection rate of $1 \times 10^{-6}$ was found at the energy deposit of $ 1\times 10^{19}$ eV %
which was proportional to square of the X-ray intensity suggesting the effect to be coherent scattering. %
The decay time of the reflection was several seconds.
This effect implies a large scale natural rock salt formation could be utilized like %
a bubble chamber irradiated by radio wave instead of visible light to detect GZK neutrinos.
\PACS{
      {61.80.Cb}{X-ray effects}   \and
      {95.55Vj}{Neutrino, pion, and other elementary particle detectors: cosmic ray detectors}
     } 
} 
\maketitle
\section{Introduction}
\label{intro}
GZK (Greisen, Zatsepin and Kuzmin) neutrinos are generated by collision between %
ultra high energy (UHE) cosmic rays ($\geq 4 \times 10^{19}$ eV) and  cosmic microwave background %
Ref.~\cite{Greisen}. %
Both have been observed, then GZK neutrinos ($\geq 10^{16}$ eV) %
have been justified to exist although the flux is very low ($\sim$ 1 km$^{-2}$ day$^{-1}$). %
The flux can be estimated to a certain extent %
since flux of UHE cosmic rays and density of cosmic microwave background are known. %
GZK neutrinos could become a standard candle of UHE neutrinos for finding out other %
conceivable UHE neutrino sources such as active galactic nuclei, topological defects and $\gamma$-ray bursts etc. %
in the universe. %
In order to detect GZK neutrinos, a huge mass of a detection medium as large as %
50 Gt (3 $\times$ 3 $\times$ 3 km$^3$ for a rock salt case) is needed due to their low flux. %
Detecting radio wave from Askryan effect (coherent Cherenkov effect) Ref.~\cite{Askaryan} is a %
promising way to utilize a large mass of natural rock salt Ref.~\cite{Stanley} or %
ice bed in Antarctica %
but it needs a lot of bore holes to be installed for radio wave detection antennas.

\begin{figure*}
\includegraphics[width=1.\textwidth,height=0.68\textwidth,angle=0]{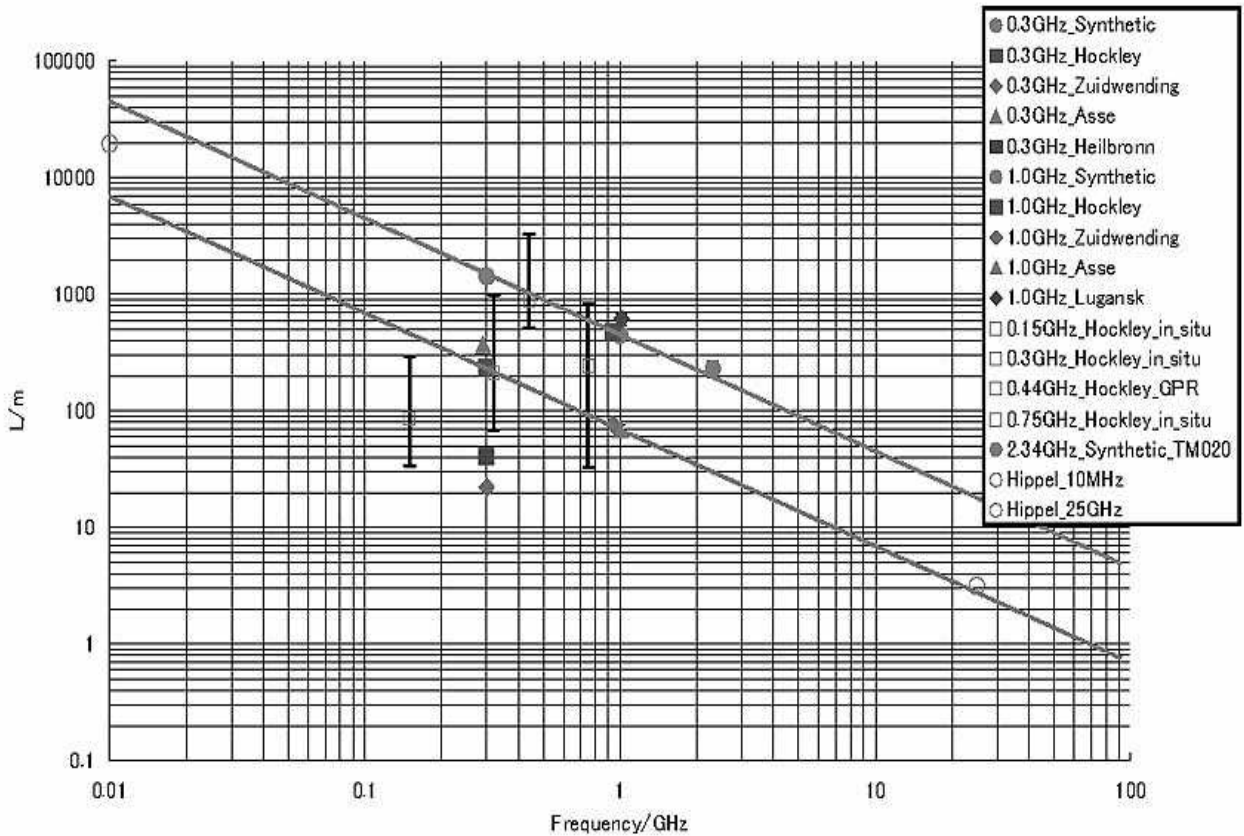}
\caption{Attenuation lengths for electric field are plotted with respect to frequency for %
synthetic and natural rock salt. %
The upper and the lower straight lines are %
fitted to synthetic and natural rock salt samples of Asse salt mine, respectively, %
assuming loss tangent is constant with respect to the frequency.}
\label{fig:attenuation}       
\end{figure*}

We had measured attenuation length of natural rock salt samples as well as %
synthetic rock salt samples of single crystal by a perturbed cavity resonator method. %
We found long attenuation lengths for electric field %
more than 200 m at 0.3 - 1 GHz in the samples of natural rock salt %
Ref.~\cite{Chiba} as shown in fig. \ref{fig:attenuation}. %
Data noted as "Hockley\_in\_situ" in the legend are %
in-situ measurements at Hockley salt mine Ref.~\cite{Hockley}.
The long attenuation length allows us to %
utilize a rock salt formation as a UHE neutrino detector Ref.~\cite{Gorham}.

%
A basic study of a new detection method has been carried out to measure %
radio wave reflection from electromagnetic energy deposit by X-ray irradiation %
in a small rock salt sample. %
If we could detect the radio wave reflection by an enough reflection rate,
a large scale natural rock salt formation would work as %
a radio bubble chamber using radio wave instead of visible light to get %
3 dimensional information of the energy deposit of the shower generated  by the UHE neutrino.

\section{Experiment}
\label{sec:2} 


We studied microwave reflection rate from a small rock salt sample irradiated by X rays, %
set in a metal wave guide. Decay time of the reflection rate after the irradiation and %
the coherency of the microwave scattering were measured. 

\subsection{X ray irradiation and apparatus}
\label{sec:2.1}




A X-ray beam of synchrotron radiation from a bending magnet %
was supplied to NE5A station by 6.5 GeV Accumulation Ring %
(AR) of KEK photon factory with the electron beam current of 40 - 60 mA. %
The X ray was a white spectrum ranging 8 - 100 keV with the number of photons of N$_\gamma = 10^{15}$ s$^{-1}$ %
at the beam current of 60 mA. In order to get a strong X-ray irradiation, we did not use a spectroscope %
in the beam line.
The X ray was chopped by 4 mm thick lead circular disk with a 4 $\times 4$ mm$^2$ orifice at the radius of 50 mm. %
The disk was rotated by a stepping motor. %
Duty of the irradiation was 1.27\% in time. %
The irradiation time was controlled by revolving speed of the disk. %
A synthetic single crystal of rock salt with a size of 2 $\times$ 2 $\times$ 10 mm$^3$ was set %
45$^{\circ}$ rotated to the square orifice sides to reduce the microwave reflection from the sample. %
It was inserted in a X band rectangular metal wave guide of $TE_{10}$ mode. %
The X ray was irradiated to the longitudinal direction of the sample and %
the energy deposit was $1\times 10^{19}$ eV s$^{-1}$ which was calculated by geant4 simulation. %
The microwave was irradiated to the sample in the lateral direction %
with the electric field being parallel to the X-ray direction of travel. %

\begin{figure}
\includegraphics[width=0.45\textwidth,height=0.3\textwidth,angle=0]{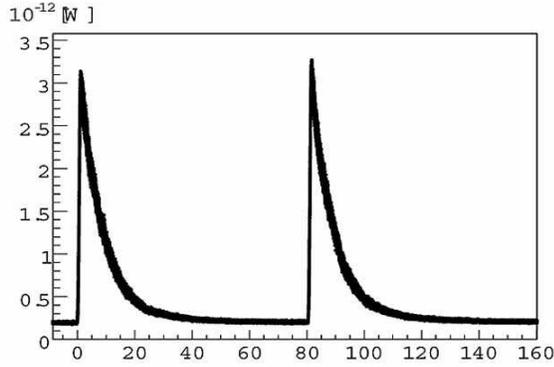}
\caption{Reflected power of the microwave against the elapse of time in a unit of s. X ray was irradiated %
1.7 s for each.}
\label{fig:reflection}       
\end{figure}

Both X ray and a continuous coherent microwave %
of 9.4 GHz with 10$^{-4}$ W from an oscillator were irradiated %
simultaneously to the rock salt to measure the reflection of microwave from %
the electromagnetic energy deposit. %
The microwave was fed to the waveguide through a coaxial-to-waveguide converter. %
The wave length of the microwave %
was 31.9 mm in vacuum and 13.3 mm in rock salt. Refractive index of rock salt is 2.4. %
Null method was employed to measure a very small reflection signal. %
Before the X-ray irradiation, the microwave reflection signal and a split signal %
from the oscillator were tuned so that the phase difference was $\pi$ radian and the amplitudes were equal. %
Consequently the sum of the reflection and the split signals became zero. %
The sum was fed to a receiver equipped with a logarithmic amplifier %
which was sensitive as low as $10^{-14}$ W. %
The output of the receiver was recorded by a digital oscilloscope as shown in fig. \ref{fig:reflection}.
%
%
\begin{figure}
\includegraphics[width=0.45\textwidth,height=0.3\textwidth,angle=0]{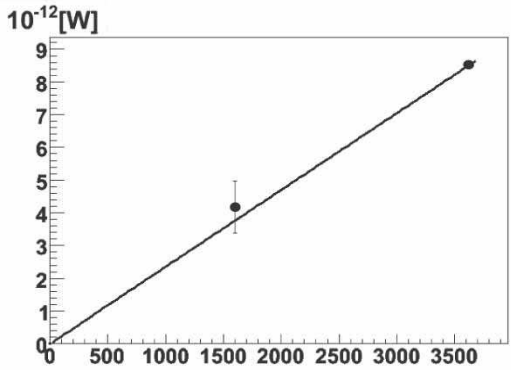}
\caption{Reflection of the microwave in power with respect to square of the AR beam current in a unit of (mA)$^2$.}
\label{fig:beam_current}       
\end{figure}
\subsection{Data analysis}
\label{sec:2.2}

The reflected power of the microwave with respect to the elapsed time from %
the X ray irradiation of 1.7 s was measured as shown in fig.\ref{fig:reflection}. 
The irradiation time of 1.7 s was a duration from a time of the forward side of the orifice %
touching the first corner %
of the rock salt sample and the backward side of the orifice crossing the last corner of the rock salt sample. %
Full open time of the orifice to the salt sample was 0.29 s during the irradiation time of 1.7 s.
Averaging the irradiation time of 1.7, 4.8, 9.6 and 16.0 s, the decay time was $8.3 \pm 1.5$ s %
at the AR beam current of 60 mA. %
The reflection rate ($\Gamma = 1 \times 10^{-6}$) with regard to power is evaluated %
for the energy deposit of $1\times 10^{19}$ eV %
taking account of loss in the waveguide circuit and the geometrical cross section %
ratio between the rock salt sample and the waveguide. %

The scattering mechanism is not known yet. But electrons, color centers, %
phonons, polarons and polaritons etc. %
could be candidates of particles as targets of the microwave scattering. %
The lifetime of the particles should be as long as several seconds. %
Energy deposit during X ray irradiation would be accumulated in the lifetime.
We study whether the scattering is coherent or not. %
Reflected power ($P_r$) is a square of electric field vector sum %
at a receiving antenna in a waveguide-to-coaxial converter as in eq. (\ref{eq:vector_sum}). %
\begin{equation}
P_r = \vert \mbox{{\boldmath $E_1$}} +  \mbox{{\boldmath $E_2$}} + \cdots + %
\mbox{\boldmath $E_i$} + \cdots + \mbox{{\boldmath $E_n$}} \vert^2 
\label{eq:vector_sum} %
\end{equation}
\begin{equation}
= \vert \mbox{\boldmath $E_1$}\vert^2 + \vert \mbox{\boldmath $E_2$}\vert^2 + \cdots +%
\vert \mbox{\boldmath $E_n$}\vert^2 + \cdots + %
2\mbox{\boldmath $E_i$} \cdot \mbox{\boldmath $E_j$} + \cdots %
\label{eq:expand} %
\end{equation}
\begin{equation}
= n^x\vert \mbox{\boldmath $E_i$} \vert^2 
\label{eq:square_sum}
\end{equation}
Where $n$ is number of the particles. %
Each vector of \mbox{\boldmath $E_i$} is electric field come from each particle. %
We expand eq. (\ref{eq:vector_sum}) and get eq. (\ref{eq:expand}) which includes the cross terms. %
If the wavelength of the microwave in the rock salt is well longer than distances between %
the particles, phase of the scattered wave is almost the same and the cross terms are %
added constructively. %
We define coherence parameter $x$ which spans from 2 to 1 for %
full coherence to null coherence, respectively. %
The null coherence means the scattered waves are in random phase. %
The scattered power is proportional to square of $n$ as written in eq. %
(\ref{eq:square_sum}) in case of the longer wavelength. 

Amount of increase in the reflected power during the full open time of 0.8 s %
were deduced in different irradiation time (beam current) %
for 16.0 s (60 mA) and, 4.8 and 9.6 s (40 mA). %
Increases of the reflected power were $(8.5 \pm 0.05)\times 10^{-12}$ and $(4.2 \pm 0.8)\times 10^{-12}$ W %
for the beam current of 60 and 40 mA, respectively, %
as shown in fig. \ref{fig:beam_current}. %
The coherence parameter was $x = 1.8 \pm 0.4$.

A sharp rise at the beginning of the irradiation in fig. \ref{fig:reflection}, we fitted the curve with %
eq. (\ref{eq:square_sum}) where $n$ is substituted to the elapse of time from %
the beginning of the irradiation. %
The number $n$ increases proportionally with the elapse of irradiation time. %
The coherence parameter became $x = 1.7\pm0.6$ which is consistent with the beam current case described above.
Both analyses show that the scattering is coherent. %
It means that each particle interacts with incident microwave independently %
and the distances among them should be considerably shorter than the wavelength of microwave %
in the rock salt sample.

Number of particles is estimated assuming Thomson scattering cross section %
of $\sigma_T = 6.653 \times 10^{-25}$ cm$^2$ between the microwave and the particles. %
We get the number of particles assuming full coherence of $x = 2$ in eq. %
(\ref{eq:particle}):
\begin{equation}
n = \sqrt{\frac{\Gamma}{\sigma_T}} = 1.2\times 10^9
\label{eq:particle}
\end{equation}
We estimate production cross section of the particles ($\sigma_t$) in the interaction between %
the X ray and electrons %
($N_e = 3\times 10^{22}$) in the rock salt sample. We get the cross section in eq. (\ref{eq:cross_section}):
\begin{equation}
\sigma_t = \frac{n}{N_\gamma N_e} = 3\times 10^{-29} \textrm{cm}^2
\label{eq:cross_section}
\end{equation}
The cross section is $10^{-4}$ smaller compared with Compton scattering cross section of %
the order of $10^{-25}$ cm$^2$.
It might mean that the particles were not free electrons produced by Compton effect and %
stopped by ionization loss inside the rock salt sample .

Most of the coherence of scattering are %
among particles in the transverse region when the wavelength are comparable to the longitudinal size 
of the electromagnetic energy deposit. %
We compare the coherence among particles in the transverse region of %
an actual UHE shower with the rock salt sample used in this experiment. %
We observed the coherent scattering with the transverse size of $w$ = 2 mm in the sample for %
the wavelength of $\lambda = 13.3$ mm microwave (9.4 GHz) in the rock salt. %
The ratio is $w \lambda^{-1}$ = 0.15. If we use the frequency of 10 MHz for %
the actual UHE electromagnetic shower keeping the same ratio of 0.15, the transverse size of 190 cm %
is allowed to a coherent scattering. %
The size seems to be large enough compared with the lateral size of the UHE shower.

We estimate the range of a radar in the rock salt formation assuming %
transmitting and receiving antennas having isotropic sensitivity. %
Peak power of the radar is assumed as 1 MW - 1 GW with a pulse duration of %
1 $\mu$s which means that spacial resolution is about 100 m %
along a direction between the antenna and a UHE shower. 
The reflection rate of $\Gamma$ is proportional to square of %
an electromagnetic energy deposit ($E$ [eV]) in the rock salt. %
We get reflection rate using %
$\Gamma = 10^{-6}$ at the energy deposit of $10^{19}$ eV as in eq. (\ref{eq:reflection}):
\begin{equation}
\Gamma = 10^{-6}\cdot (10^{19}\cdot E)^2
\label{eq:reflection}
\end{equation}
Attenuation rate $\alpha$ in power is calculated in eq. (\ref{eq:attenuation}):
\begin{equation}
\alpha = (\exp (-2R/L))^2
\label{eq:attenuation}
\end{equation}
Where \textit{R} is a range of the radar between the antenna and the UHE shower, %
and \textit{L} is an attenuation length %
of radio wave for electric field.
Receiving power is expressed by eq. (\ref{eq:radar}):
\begin{equation}
P_r = P_e \cdot \alpha \cdot \Gamma \cdot (4\pi R^2)^{-2} \cdot S_1 \cdot S_2
\label{eq:radar}
\end{equation}
Where $P_e$ is an emission power at the transmission antenna, $S_1$ and $S_2$ are effective %
cross sections of the UHE shower and the receiving antenna, respectively.
From our measurements of the perturbed cavity resonator method as in fig. \ref{fig:attenuation}, %
the attenuation length is extrapolated to 10 MHz. Then we get $L = 7000$ m.
We assume that the detection limit of radio wave is $P_r \geq 10^{-14}$ W and %
the effective cross sections of the UHE shower and the receiving antenna are %
$S_1 = S_2 = 1$ m$^2$.
We get the range of the radar as shown in table 1.

\begin{table}
\caption{Range of radar (peak power of 1 MW \& 1 GW) against the shower energy deposit and %
reflection rate $\Gamma$.}
\label{tab:1}       
\begin{tabular}{llll}
\hline\noalign{\smallskip}
shower \textit{E}/eV & $\Gamma$ & \textit{R}/m(1 MW) & \textit{R}/m(1 GW) \\
\noalign{\smallskip}\hline\noalign{\smallskip}
$10^{17}$ & $10^{-10}$ & 88  & 469 \\
$10^{18}$ & $10^{-8}$ & 271  & 1310 \\
$10^{19}$ & $10^{-6}$ & 796  & 3180 \\
\noalign{\smallskip}\hline
\end{tabular}

\end{table}
\section{Summary}
\label{sec:3} 
Microwave was reflected at the rate of $10^{-6}$ from X ray irradiated rock salt %
in the energy deposit of $1\times 10^{19}$ eV. %
Life time of particles which was responsible to the decay time of %
the microwave reflection was several seconds. %
It is long enough to employ periodic transmission of radar pulses without triggered by %
reception of Askaryan radio wave. %
Microwave scattering from the irradiated rock salt was coherent %
but the particle species working as the scattering targets is not known. %

Power of radio wave emitted by Askaryan effect   %
increases proportionally to the frequency. %
At the higher frequency the attenuation length becomes short %
as shown in fig. \ref{fig:attenuation}. %
On the contrary radar method is not imposed such a restriction on the frequency. %
Strong artificial radar pulses are available for the radar method. %
Consequently, we could get long range of detection. %

Times and amplitude from several receiving antennas could give us 3 dimensional %
information of the UHE shower with its energy. %
In order to utilize a salt dome with a diameter of 3 km and a depth of 3 km (50 Gt), %
several bore holes are needed in which the transmitting and receiving antennas are installed. 
Radar method would have a potential to realize Salt Neutrino Detector to act like a Radio %
Bubble Chamber to detect GZK neutrinos. %

If we could get the peak power of the radar considerably larger than 1 GW, %
the range becomes long enough to know whether GZK neutrinos exist or not without expensive bore holes. %
The antennas would be installed slightly under a floor in an excavated space of a rock salt dome. %

\begin{acknowledgement}
Work is partially supported by a Grant in Aid for Scientific Research for Ministry of Education, Science, %
Technology and Sports and Culture of Japan, and Funds of Tokubetsu Kenkyuhi, at Seikei University. %
We express deep appreciation to KEK-PF staffs, especially Dr. Kazuyuki Hyodo who extended us his hospitality %
throughout this experiment, without him this measurement could not be carried out smoothly.
\end{acknowledgement}
%

%

\begin{thebibliography}{999}
%
%

\bibitem{Greisen}
Greisen K, Phys. Rev. Lett. \textbf{16}, (1966) 748;%
Zatsepin G T and Kuz'min V A, Zh. Eksp. \& Teor. Fiz., Pis' ma Red.\textbf{4}, (1966) 114%
(Sov. Phys.-JETP Lett.\textbf{4}, (1966) 78).
%
\bibitem{Askaryan}
Askaryan G A, Zh. Eksp. \& Teor. Fiz. \textbf{41}, (1961) 616 ( Phys. JETP. \textbf{14}, (1962) 441);%
Askaryan G A, Phys. JETP. \textbf{48}, (1961) 988 (\textbf{21}, (1965) 658; %
Zas E, Halzen F and Stanev T, Phys. Rev. D \textbf{45}, (1992) 362.
%
\bibitem{Stanley}
Stanley J L, \textit{Handbook of World Salt Resources} (New York: Plenum Press);%
T. H. Michel, \textit{Salt Domes} (Houston : Gulf Publishing Company)
%
\bibitem{Chiba}
Chiba M \textit{et al}., \textit{Proc. 1st Int. Workshop for Radio Detection of High Energy Particles} (AIP Conf. Proc. vol 579, 2000) 204; %
Chiba M \textit{et al}., \textit{Proc. of the First NCTS Workshop Astroparticle Physics} (World Scientific Publishing Co. Ltd. 2002) 99; %
Kamijo T and Chiba M, \textit{Proc. of SPIE Particle Astrophysics Instrumentation} (SPIE vol 4858 Bellingham WA 2003) 151; %
Chiba M \textit{et al}. \textit{Proc. of the International Workshop, ARENA2005} (World Scientific Publishing Co. Ltd. 2006) 50; %
Watanabe Y \textit{et al}. \textit{ibid.} 25; %
Watanabe Y \textit{et al}. \textit{Proc. of the International Symposium on Origin of Matter and Evolution of Galaxies Vol 847} (AIP conference proceedings) 491. %
%
\bibitem{Gorham}
Chiba M \textit{et al}., Physics of Atomic Nuclei \textbf{67}, (2004) 2050; %
Gorham P \textit{et al}., Phys. Rev. D \textbf{72}, (2005) 023002; %
Saltzberg D \textit{et al}., \textit{Proc. of SPIE Particle Astrophysics Instrumentation} (SPIE vol 4858 Bellingham WA 2003) 191.
%
\bibitem{Hockley}
Gorham P \textit{et al}., Nucl. Instrum. \& Methods A \textbf{490}, (2002) 476.


\end{thebibliography}
%

\end{document}